\documentclass[12pt]{iopart}

\usepackage{graphicx}
\usepackage{gensymb}

\begin{document}

\title{Integrated coherent matter wave circuits}

\author{C. Ryu and M. G. Boshier}

\address{Physics Division, Los Alamos National Laboratory, Los Alamos, NM 87545, USA}

\ead{boshier@lanl.gov}

\begin{abstract}
An integrated coherent matter wave circuit is a single device, analogous to an integrated optical circuit, in which coherent de Broglie waves are created and then launched into waveguides where they can be switched, divided, recombined, and detected as they propagate. Applications of such circuits include guided atom interferometers, atomtronic circuits, and precisely controlled delivery of atoms.  Here we report experiments demonstrating integrated circuits for guided coherent matter waves. The circuit elements are created with the painted potential technique, a form of time-averaged optical dipole potential in which a rapidly-moving, tightly-focused laser beam exerts forces on atoms through their electric polarizability. The source of coherent matter waves is a Bose-Einstein condensate (BEC).  We launch BECs into painted waveguides that guide them around bends and form switches, phase coherent beamsplitters, and closed circuits.   These are the basic elements that are needed to engineer arbitrarily complex matter wave circuitry.
\end{abstract}


\maketitle

\section{Introduction}
It has been a longstanding goal in the field of atom optics to realize an integrated coherent matter wave circuit \cite{Dekker2000, Schneble2003, Andersson2002, Godun2001, Fortagh2005}.  This concept envisions a matter wave analog of an integrated optical circuit: a single device in which coherent de Broglie waves would be created and then launched into waveguides where they can be switched, divided, recombined, and detected as they propagate.  Research to develop coherent matter wave circuits is motivated in part by the many potential applications of this technology.  One important aim is the creation of waveguide atom interferometers \cite{Cronin2009}, which have applications ranging from fundamental physics to various forms of sensing.  For example, when the interferometer splits the moving matter waves into two wavepackets and then recombines them after the separated wavepackets have traveled along different paths that enclose an area, the device will be sensitive to rotations through the Sagnac phase \cite{Cronin2009}.  A ring waveguide geometry should allow for a large enclosed area relative to the size of the device and also permit making many round trips in the interferometer.  These are considerable advantages over free space atom gyroscopes \cite{Gustavson1997}.  A second circuit application lies in the emerging field of atomtronics \cite{Seaman2007, Pepino2010}, which develops cold atom analogs of electronic devices such as diodes \cite{Ruschhaupt2004}, transistors \cite{Stickney2007, Pepino2009, Micheli2004, Vaishnav2008}, and batteries \cite{Zozulya2013}.  As with standard electronics, one wishes to connect these building block devices together to create complex circuits with the desired functionality for the current of atoms.  Other circuit applications include precisely delivering atoms to a desired point in space with a specified velocity, realizing a kind of "atom laser pointer" that may be useful in technologies such as atom lithography.  In this paper we describe experiments that create simple circuits for propagating coherent atomic matter waves.  These circuits demonstrate the basic elements that are needed to engineer arbitrarily complex matter wave circuitry.

The problem of realizing the matter wave equivalent to the laser source of integrated optics, namely creating coherent de Broglie waves in a waveguide, was solved in 1995 with the development of techniques to form an atomic Bose-Einstein condensate (BEC) \cite{Anderson1995, Davis1995, Bradley1995}.  In the past fifteen years several of the other elements required to construct a matter wave circuit have been demonstrated individually for incoherent cold atoms that fill many spatial modes of the confining waveguide. However, extending these techniques to operate in a single spatial mode with coherent matter waves from a BEC and then incorporating the elements into integrated circuits have both proven to be difficult.   We show here how to accomplish both of these goals.

\begin{figure}
\begin{center}
\includegraphics[width=15cm]{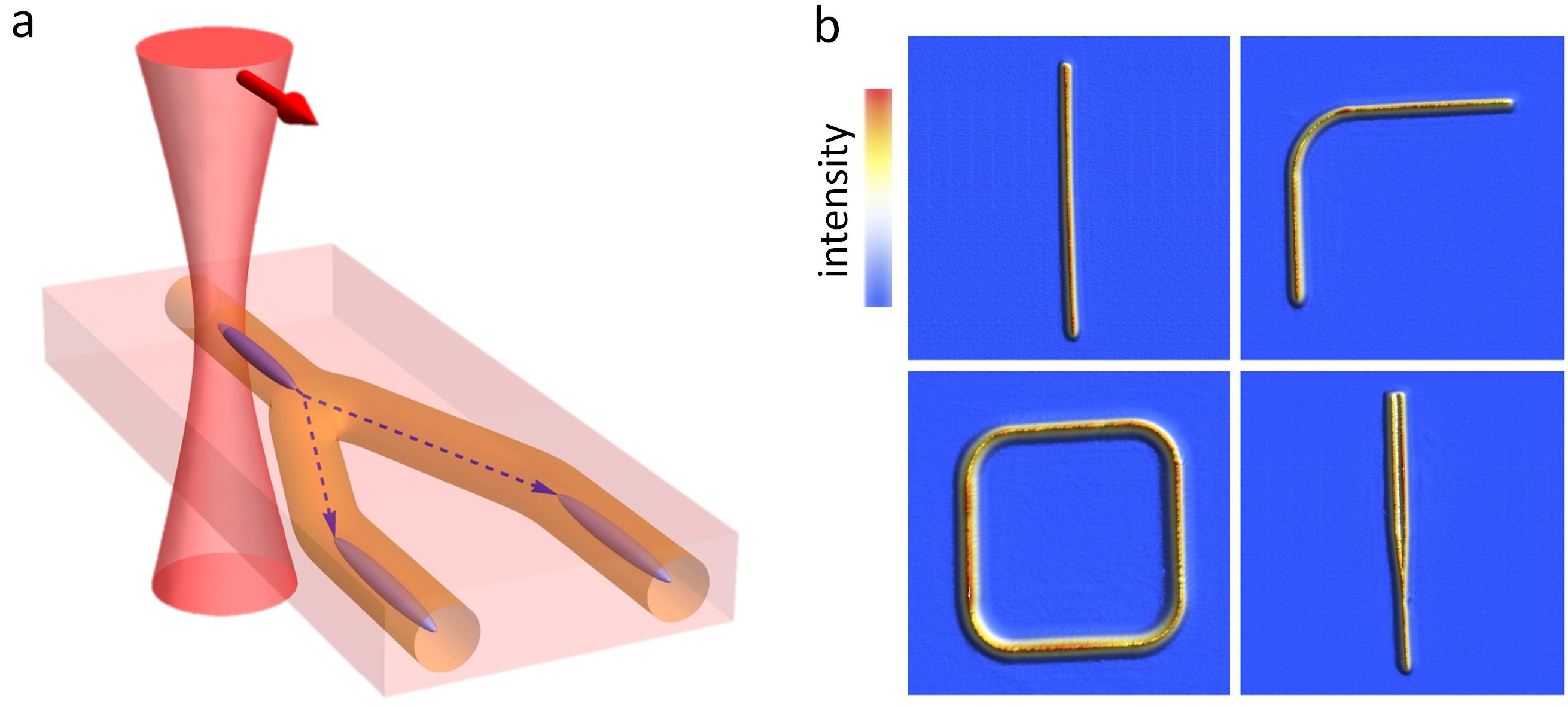}
\caption{\label{apparatus} (a)  The coherent matter waves of a moving Bose-Einstein condensate propagate along waveguide-shaped time-averaged optical dipole potentials formed by the combination of a horizontal laser light sheet and a rapidly-moving, tightly-focused vertical laser beam.  The vertical beam paints the desired waveguide geometry, here a Y-junction.  (b) Measured time-averaged laser intensity distributions used in the experiments reported here, recorded by imaging the laser intensity at the plane of the circuit onto a camera.  The intensity is rendered into three dimensions using the colour scale shown.  Clockwise from top left, with image dimensions in parentheses: straight waveguide (114\,$\mu$m\,$\times\,$114\,$\mu$m), straight waveguides connected by a circular bend (93\,$\mu$m\,$\times\,$93\,$\mu$m), Y-junction (114\,$\mu$m\,$\times\,$114\,$\mu$m), and square waveguide circuit (62\,$\mu$m\,$\times\,$62\,$\mu$m).}
\end{center}
\end{figure}

Cold atoms can be manipulated using either the magnetic Stern-Gerlach force or the optical dipole force.  Using the magnetic approach, currents in suitably-shaped conductors have been used to guide incoherent laser-cooled atoms in straight lines \cite{Key2000, Fortagh2000}, around bends \cite{Muller1999, Luo2004}, and to form Y-junction splitters for such atoms \cite{Muller2000, Cassettari2000}.  Also, BECs have been launched in toroidal \cite{Gupta2005, Arnold2006} and linear \cite{Garcia2006} guides produced with electromagnets.  The development of atom chips for cold atoms \cite{Folman2000, Dekker2000} and BEC \cite{Hansel2001} opened up possibilities for integration \cite{Berrada2013}, complex geometries, and miniaturization.  However, while coherent beamsplitters for stationary BECs have been realized on atom chips \cite{Schumm2005}, to date all propagation on atom chips has been restricted to linear guides \cite{Leanhardt2002, Fortagh2004, Wang2005}.  The optical dipole force of laser light propagating inside a hollow-core optical fiber was used in the first demonstration of atom guiding \cite{Renn1995}.  Subsequently, incoherent cold atoms from a magneto-optical trap were propagated along a miniature planar optical dipole potential waveguide above a surface \cite{Schneble2003}.  Propagation of coherent matter waves along a linear optical dipole guide formed by a collimated laser beam has been demonstrated \cite{Guerin2006, McDonald2013}, and overlapping laser beams have been used to create beamsplitters for guided cold atoms \cite{Houde2000} and for an atom laser \cite{Gattobigio2012}.  Micro-optics have been used to create beamsplitter and interferometer optical dipole potentials for cold atoms \cite{Dumke2002}.  In the limit where the moving BEC completely fills the waveguide, superfluid flow has been observed in toroidal optical dipole potentials \cite{Ryu2007, Ramanathan2011, Moulder2012, Ryu2014} and atom-SQUID devices have been demonstrated \cite{Wright2013, Ryu2013, Eckel2014}.  Finally, digital holography has created complex optical dipole potentials that might realize an atomtronic OR-gate once loading of cold atoms into the potential has been demonstrated \cite{Gaunt2012}.   However, none of the experiments discussed above has demonstrated phase-coherent splitting of propagating matter waves or single-mode matter wave propagation in waveguide sections connected by bends - two essential ingredients of a coherent matter wave circuit.  Both are realized in our experiment.

Our circuit elements are created with the painted potential technique \cite{Henderson2009}, a form of time-averaged optical dipole potential in which a rapidly-moving, tightly-focused laser beam superimposed on a sheet of laser light exerts confining forces on atoms through their electric polarizability.  Our device is analogous to an integrated optical circuit with the roles of matter and light being reversed:  while the optical circuit uses matter to guide light, here we use light to guide matter.  The matter wave source, analogous to the laser, is a BEC.   The painted potential is used to draw waveguides and waveguide structures [Fig.\,\ref{apparatus}(a)], such as bends and Y-junctions [Fig.\,\ref{apparatus}(b)].  While the proof-of-principle circuits we present here are simple, the system should be able to create any planar circuit topology that can be represented with the approximately $100 \times 100$ resolvable spots of the two-axis acousto-optic deflector that scans the painting beam \cite{Henderson2009}.  Further, because the painted potential is dynamic the matter wave circuit can be modified arbitrarily as atoms propagate through it, a degree of flexibility that is not available in atom chips.  In the experiments reported here, we launch BECs into painted waveguides that guide coherent matter waves around bends and form switches, phase coherent beamsplitters, and closed circuits.  In the following sections we discuss the implementation and performance of each of these circuit elements.

\section{Results}

\subsection{Apparatus}
Our original painted potential apparatus used for creating and manipulating BECs in arbitrary shapes has been described in detail elsewhere \cite{Henderson2009}.  Briefly, a cloud of cold $^{87}$Rb atoms is formed in a standard double magneto-optical trapping system.  After the cloud is compressed in a magnetic quadrupole trap, the painted potential overlapping the cloud is switched on.  In the configuration used here and illustrated schematically in Fig.\,\ref{apparatus}(a), the painted potential is formed from a combination of a horizontal light sheet with trapping frequency 433\,Hz and a time-averaged waveguide potential of trapping frequency 1705\,Hz and depth $1.39\,\mu$K painted by a beam with waist size $2.15\,\mu$m at repetition frequency of 10\,kHz.  Reducing the intensity of the light sheet allows relatively hot atoms to evaporate from the trap until a condensate is formed in the painted potential.  The significant improvements to the flatness and resolution of the potential described in Refs. \cite{Ryu2014, Ryu2013} are essential for realizing the experiments reported here.  To permit slow-moving wavepackets to propagate without reflection from imperfections in the waveguide potential it is necessary to flatten the time-averaged potential to better than $10\%$ of its depth.  This is achieved by changing the intensity of the painting beam appropriately as it was scanned.  Such smoothing of the potential is not possible with magnetic potentials produced by currents flowing in conductors on atom chips, which may explain why propagating matter wave circuits have not been realized with that technology.   We measure the potential flatness either by observing density variations in a BEC created to fill the entire circuit or by observing the propagation dynamics of slowly-moving condensates.  The uncorrected intensity distributions in Fig.\,\ref{apparatus}(b) show the level of smoothness present in the time-averaged potential before this flattening process is applied.

All BEC images presented here are taken in absorption, so each image corresponds to a different experimental run.  Except where a period of free expansion before imaging is specified, all images depict the BEC moving in the trapping potential.   In all cases condensate density increases from black to white, with blue corresponding to background (regions of zero density).

\subsection{BEC Launching}
\begin{figure}
\begin{center}
\includegraphics[width=14cm]{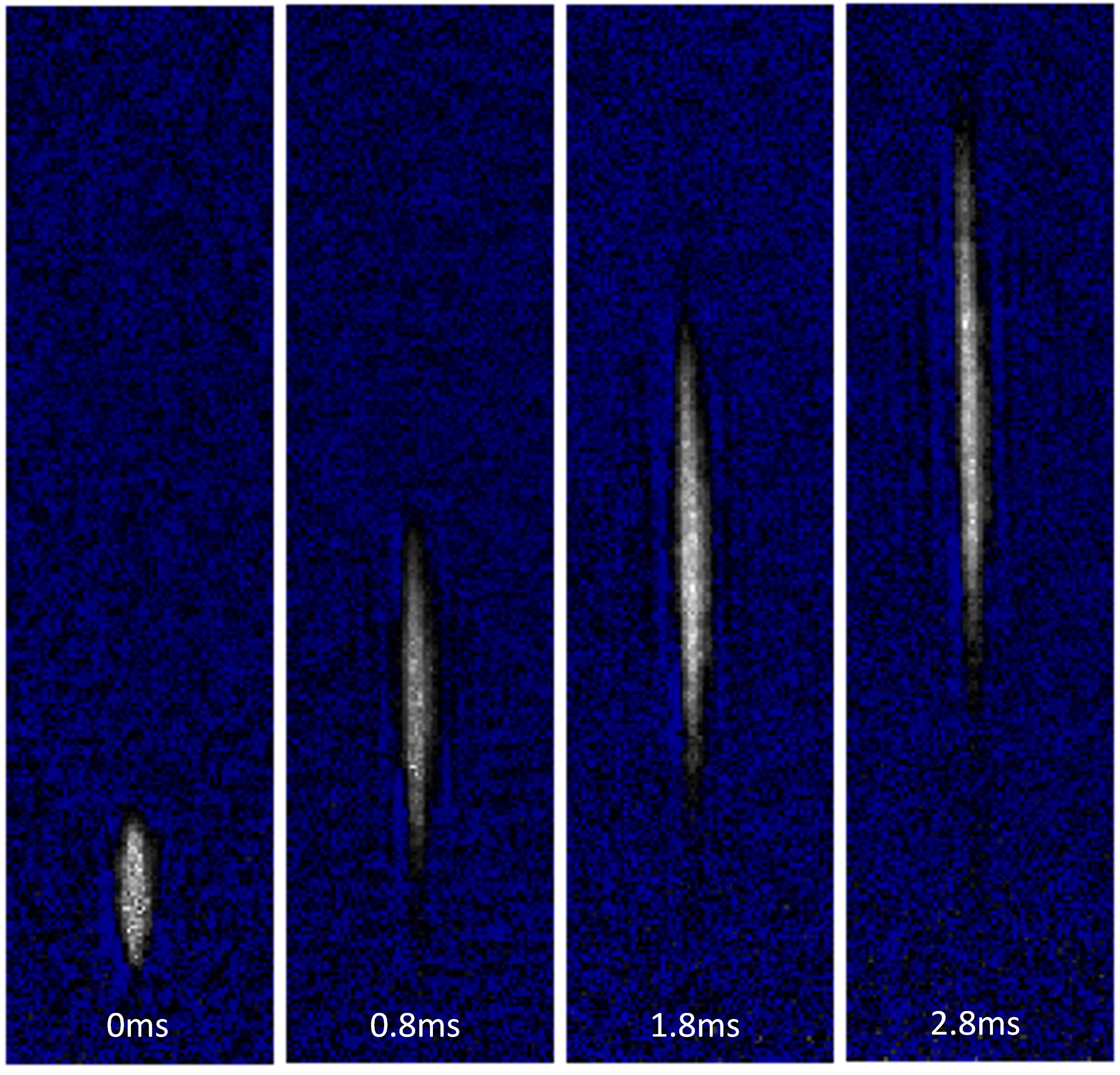}
\caption{\label{launch} A BEC launched with speed 19\,mm/s into a straight waveguide potential similar to that shown in the top left panel of Fig.\,\ref{apparatus}(b).  The launch duration was 1.3\,ms.  The times marked on each image are relative to the end of the launching process.  The dimensions of each panel are $120\,\mu$m\,$\times\,30\,\mu$m.}
\end{center}
\end{figure}

The experiments begin with the creation of a $^{87}$Rb BEC in a short ($12.4\,\mu$m long) painted waveguide.  The BEC typically contains 4000 atoms in the $|1,-1 \rangle$ state, resulting in a chemical potential of approximately $150\,$nK. The potential is then switched to the circuit geometry under investigation.  The BEC is launched with a desired velocity by spatially modulating the intensity of the painting beam to create a linear slope in the waveguide potential for the BEC to accelerate down [Fig.\,\ref{launch}].  In this way we create velocities up to 25\,mm/s.  This simple launching method has a side effect of modulating the transverse trapping potential, because it also depends on the laser intensity.  The result is excitation of a small breathing oscillation in the BEC.  Numerical simulations with the Gross-Pitaevskii equation (GPE) show, and experiment confirms, that the amplitude of the excitation is reduced to an unimportant level by making the potential slope more gentle and the acceleration time correspondingly longer (typically 1.5\,ms).  This effect could be avoided in future work by painting a transversely-uniform inclined plane potential on top of the waveguide potential.  We note that the axial expansion of the propagating wave packets seen in Fig.\,\ref{launch} is mostly due to the repulsive interactions between the atoms of the condensate.  It would be greatly reduced, and the signal to noise ratio correspondingly improved, in a BEC of $^{39}$K, where the interaction strength can be tuned through zero with a magnetic field that controls a Feshbach resonance \cite{Roati2007}.

\subsection{Waveguide Bends}
\begin{figure}
\begin{center}
\includegraphics[width=14cm]{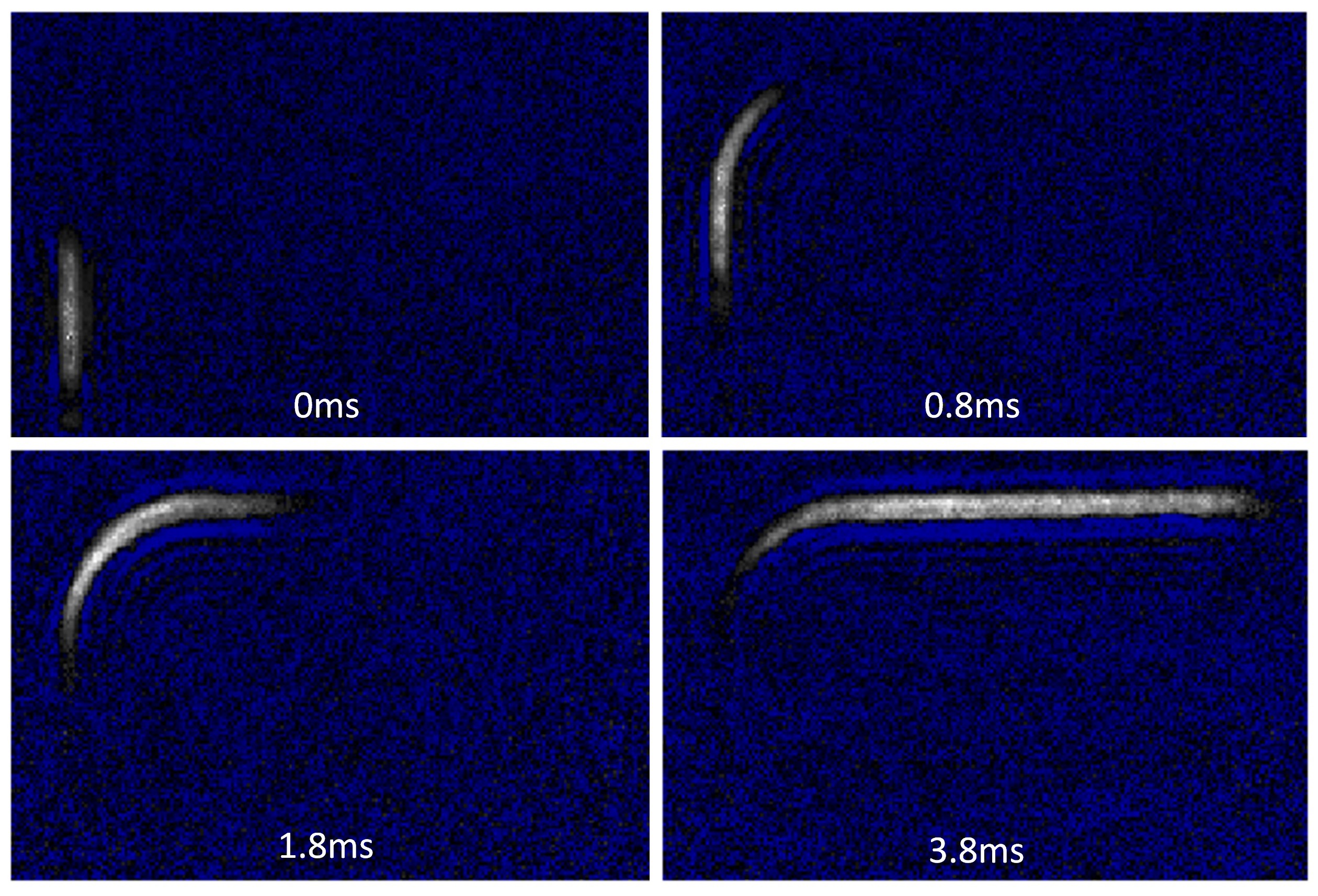}
\caption{\label{bend} A BEC propagating with speed 19\,mm/s around a $90\degree$ bend with radius $18.6\,\mu$m, a potential similar to that shown in the top right panel of Fig.\,\ref{apparatus}(b).   The launch duration was 1.3\,ms.   The times marked on each image are relative to the end of the launching process.    The dimensions of each panel are $90\,\mu$m\,$\times\,60\,\mu$m.}
\end{center}
\end{figure}

A bend connecting two straight waveguides is an essential matter wave circuit element.  While incoherent cold atoms filling hundreds of modes of a macroscopic waveguide have been propagated around bends in a waveguide \cite{Muller1999, Luo2004}, there has been no previous demonstration of atoms kept in the ground state of a transverse confining potential while propagating around bends connecting straight waveguide sections.  Fig.\,\ref{bend} shows the propagation of a BEC around such a circuit element, realizing this goal.

\begin{figure}
\includegraphics[width=16cm]{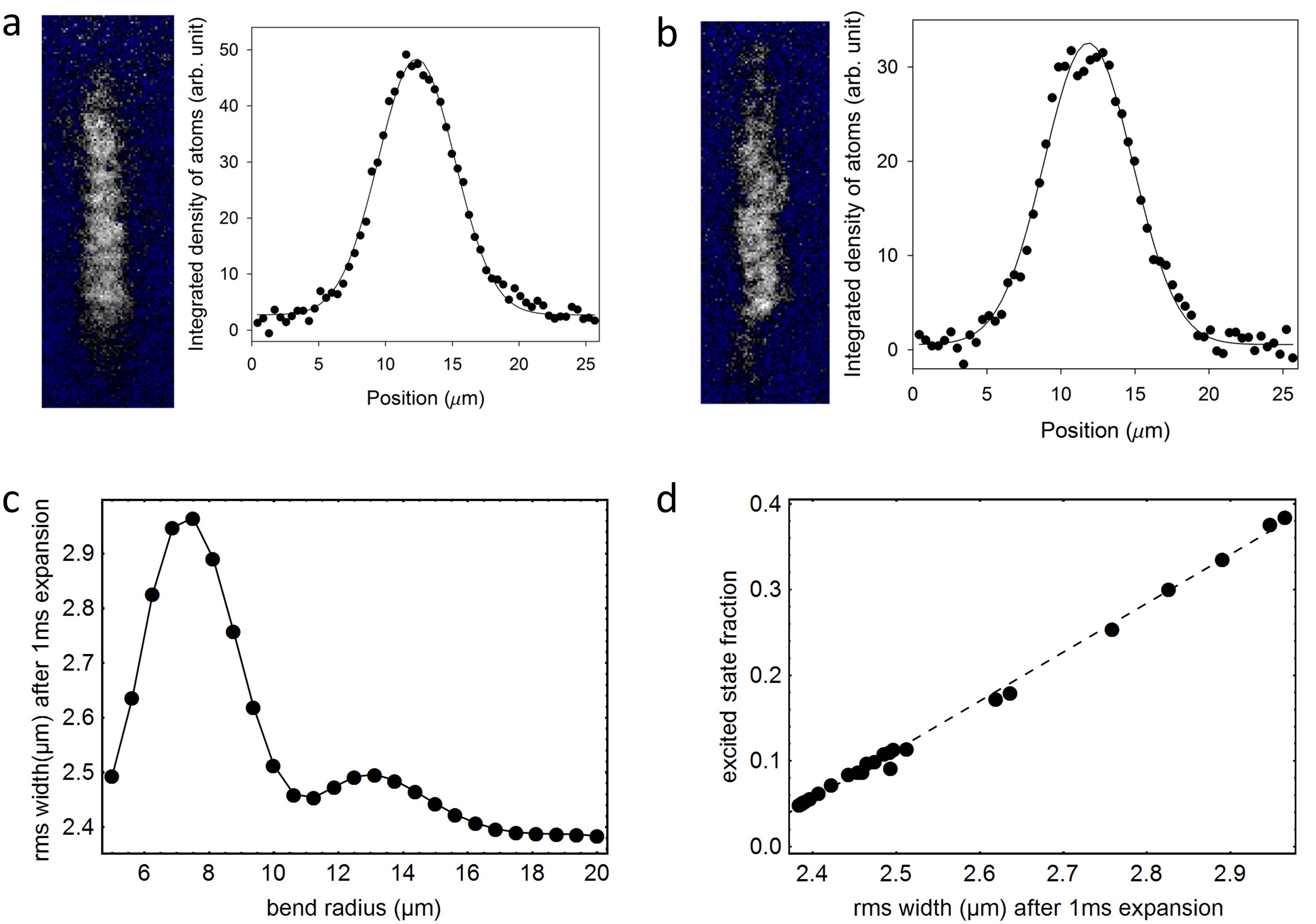}
\caption{\label{bendAnalysis} The increase in the width of a BEC allowed to expand before and after propagating around a bend can be used to determine the occupation of the ground state after the bend.  (a) Left: A BEC allowed to expand freely for 1\,ms inside the light sheet of the painted potential after propagating at speed 19\,mm/s inside a straight guide for 2.5\,ms.  Image dimensions are $77\,\mu$m\,$\times\,26\,\mu$m. Right: gaussian fit to density integrated along axis of condensate.   (b) Left: A BEC allowed to expand freely for 1\,ms inside the light sheet of the painted potential after propagating at speed 19\,mm/s inside a guide containing a bend of radius $18.6\,\mu$m for 3.3\,ms, sufficient time for the BEC to propagate past the bend.  Image dimensions are $77\,\mu$m\,$\times\,26\,\mu$m. Right: gaussian fit to density integrated along axis of condensate.  (c) Results from GPE simulations of propagation of a BEC with speed 14\,mm/s around bends of different radii followed by free expansion inside the light sheet.  The disks show the rms width after 1\,ms of free expansion inside the light sheet versus bend radius. The solid line connects the points as a guide to the eye. (d)  The disks show the condensate excited state fraction after the bend versus the rms width after 1\,ms of free expansion inside the light sheet for the simulation results in (c).  The dashed line is a linear fit to the result.  The slope of this line is used in the analysis to connect the measured increase in rms width with the decrease in ground state occupation.}
\end{figure}

An important question for many applications is: to what extent is the BEC in the ground state of the waveguide after it emerges from the bend?  To answer this question we image the BEC after switching off the waveguide potential and allowing the condensate to expand freely inside the light sheet for 1\,ms.  Fig.\,\ref{bendAnalysis}(a) shows the result obtained before the BEC enters the bend.  The same type of measurement made after the bend [Fig.\,\ref{bendAnalysis}(b)] contains an axial "wiggle".  This structure indicates that negotiating the bend creates a collective excitation of the condensate (not a thermal gas of non-condensed atoms).  Numerical simulations with the GPE of the propagation of BECs around bends exhibit such coherent excitations.  The simulations also show that the root mean square (rms) width of the condensate observed after the free expansion is a good measure of the degree of excitation (Fig.\ref{bendAnalysis}).  Specifically, if $\psi(r)$ is a normalized transverse wavefunction of the ground state in the guide, we compute for the normalized GPE solution wavefunction after the bend, $\Psi(r,z)$, the axially integrated projection, $p_{0}$, of $\Psi$ onto $\psi$:
\begin{equation}
 p_{0}=\int  dz | \int \psi(r) \Psi(r,z) dr |^{2}.
\end{equation}
The GPE simulations show that a BEC propagated through a range of bend radii exhibits varying degrees of excitation after the bend, and hence varying rms widths when the BEC is allowed to expand after passing around the bend [Fig.\,\ref{bendAnalysis}(c)].  For each simulation we compute as above the excited state fraction in the waveguide, $1-p_{0}$, after the bend.  We find that it is an approximately linear function of the rms width $\sigma$ with slope $d (1-p_{0})/d \sigma = 0.6/\mu$m [Fig.\ref{bendAnalysis}(d)].  For a condensate containing 4000\,atoms the width before the bend [Fig.\,\ref{bendAnalysis}(a)] is $\sigma = 2.68(15)\,\mu$m, and the width after the bend [Fig.\,\ref{bendAnalysis}(b)], corrected for the slightly longer propagation time, is $\sigma = 2.82(6)\,\mu$m.  From the difference, 0.14(16)\,$\mu$m, we thus infer that the relative occupation of the ground state after the bend is 0.92(9).  While this represents a relatively small degree of excitation (less than $8\%$ per bend), a significantly lower amount of excitation would be necessary if the matter waves are to negotiate many bends without substantial excitation.  As we now discuss, there are a number of strategies available to achieve this goal.

Theoretical studies of the excitation of a wave packet propagating around a circular bend \cite{Bromley2004} have found that the degree of transverse excitation depends sensitively on an interplay between bend radius and BEC velocity, with the excitation being minimal when the transverse oscillation of the condensate in the curved guide returns it to the center of the waveguide just as it exits the bend [this effect is responsible for the oscillatory behavior seen in Fig.\,\ref{bendAnalysis}(c)].  While this prediction could be studied in future experiments, there may be better solutions to reducing excitation than fine tuning the BEC velocity for the particular bend radius.  An analogous problem arises in optical waveguides, where it is known that introducing an offset between the center of the bend and the center of the straight section improves coupling into the lowest mode of the curved guide \cite{Kitoh1995}.   In the matter wave circuit implementation of this idea, the bend radius would be decreased slightly so that the BEC enters the bend waveguide off-center by just the distance that allows the transverse waveguide potential to provide the required classical centripetal force.  Alternatively, it should be possible to reduce excitations if the transition between straight waveguide and bend is made more gradual than it is for the simple circular bend employed above, where there is a discontinuous step increase in curvature at the start of the bend.  One possibility is the clothoid (Euler spiral) employed in highway and railway transition curves, which has the useful property that the curvature is a linear function of arc length along the curve.  Numerical simulations confirm that these techniques significantly reduce excitation out of the ground state at bends \cite{Boshier2014}.  The painted potential should be able to realize both the offset and clothoid approaches.

\subsection{Closed Waveguide Circuit}

\begin{figure}
\begin{center}
\includegraphics[width=14cm]{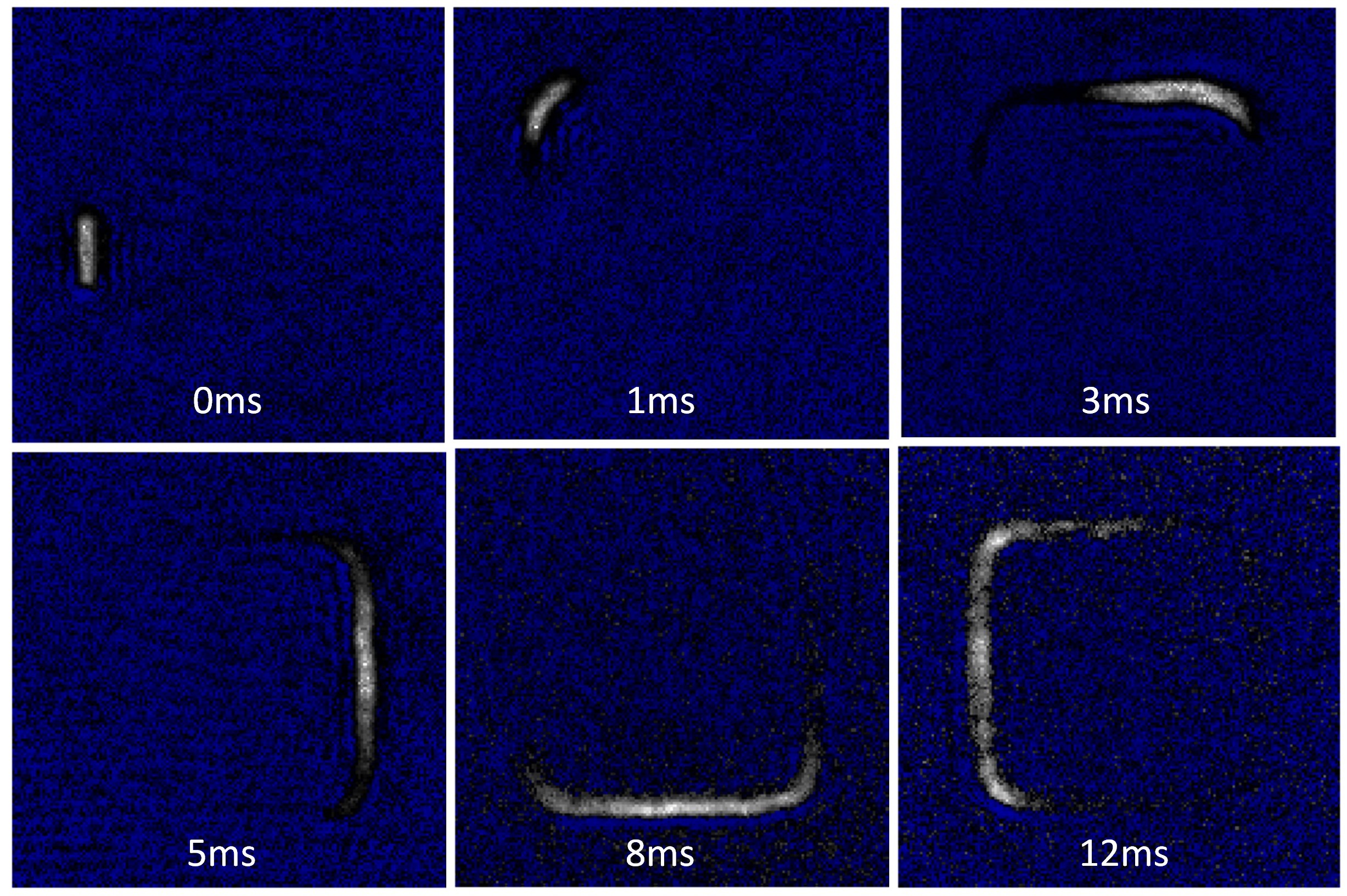}
\caption{\label{stadium} A BEC launched with speed 19\,mm/s into a closed waveguide formed from straight sections and four $90\degree$ bends of radius $9.3\,\mu$m, forming a potential similar to that shown in the bottom left panel of Fig.\,\ref{apparatus}(b).  The launch duration was 0.8\,ms.  The times marked on each image are relative to the end of the launching process.  The dimensions of each image are $70\,\mu$m$\,\times\,70\,\mu$m.}
\end{center}
\end{figure}

Figure\,\ref{stadium} shows bends and straight sections connected together to form a closed waveguide circuit, demonstrating that several non-trivial circuit elements can be combined and function well.  Such a circuit might be used, for example, to realize a Sagnac interferometer by using Bragg diffraction to coherently split a BEC in a straight section into wave packets counter-propagating around the circuit \cite{Wang2005, McDonald2013}, and then recombining them after a number of complete circuits to measure the Sagnac phase.

\subsection{Waveguide Junction}

\begin{figure}[!]
\begin{center}
\includegraphics[width=14cm]{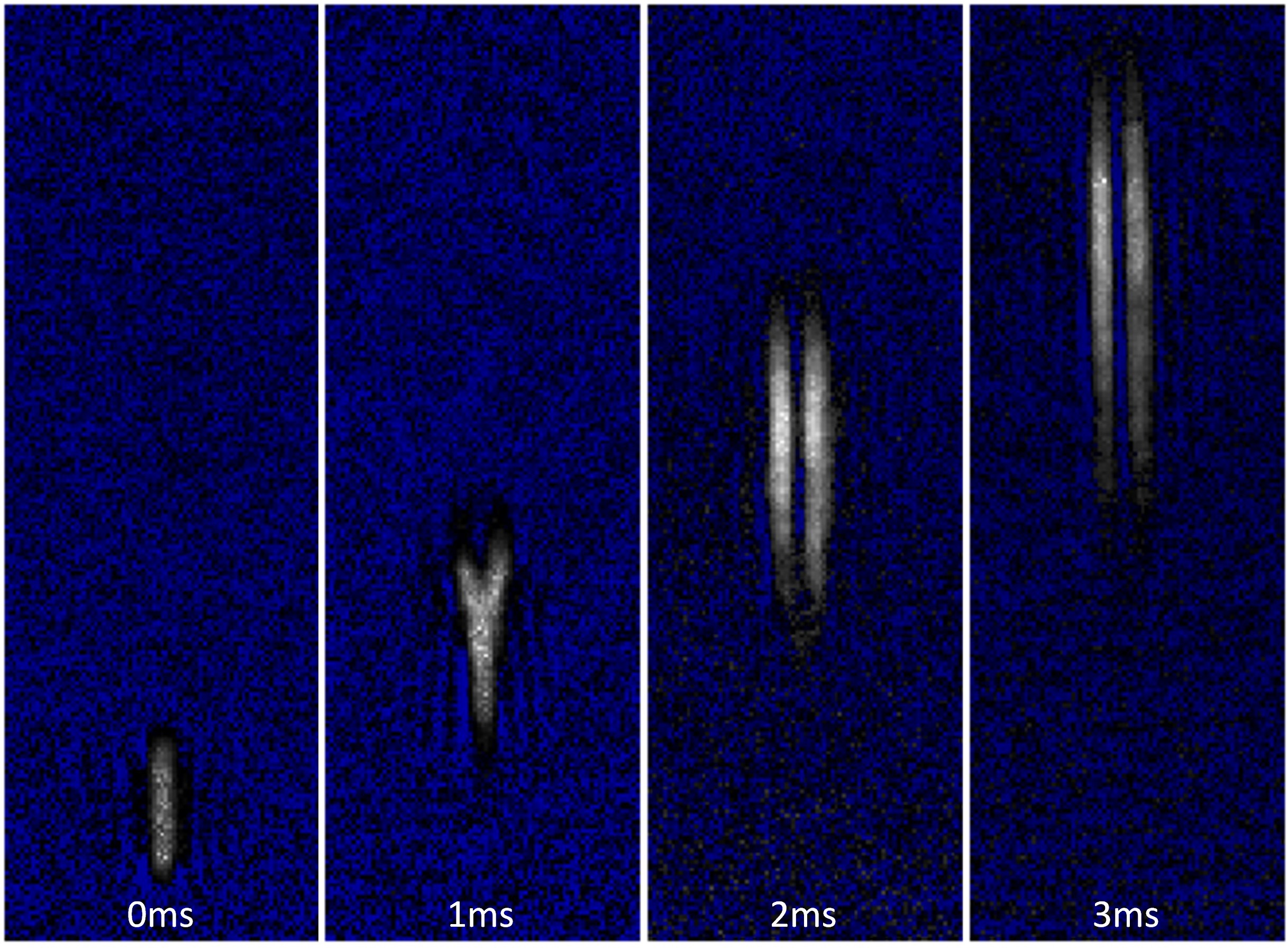}
\caption{\label{Yjunction}  A BEC propagating through a Y-junction at speed 21\,mm/s.  The arm separation is $3.7\,\mu$m, forming a potential  similar to that shown in the bottom right panel of Fig.\,\ref{apparatus}(b).  The dimensions of each image are 100\,$\mu$m\,$\times\,34\,\mu$m.  The propagation times on each panel are relative to the end of the launching process.}
\end{center}
\end{figure}

\begin{figure}[!]
\begin{center}
\includegraphics[width=14cm]{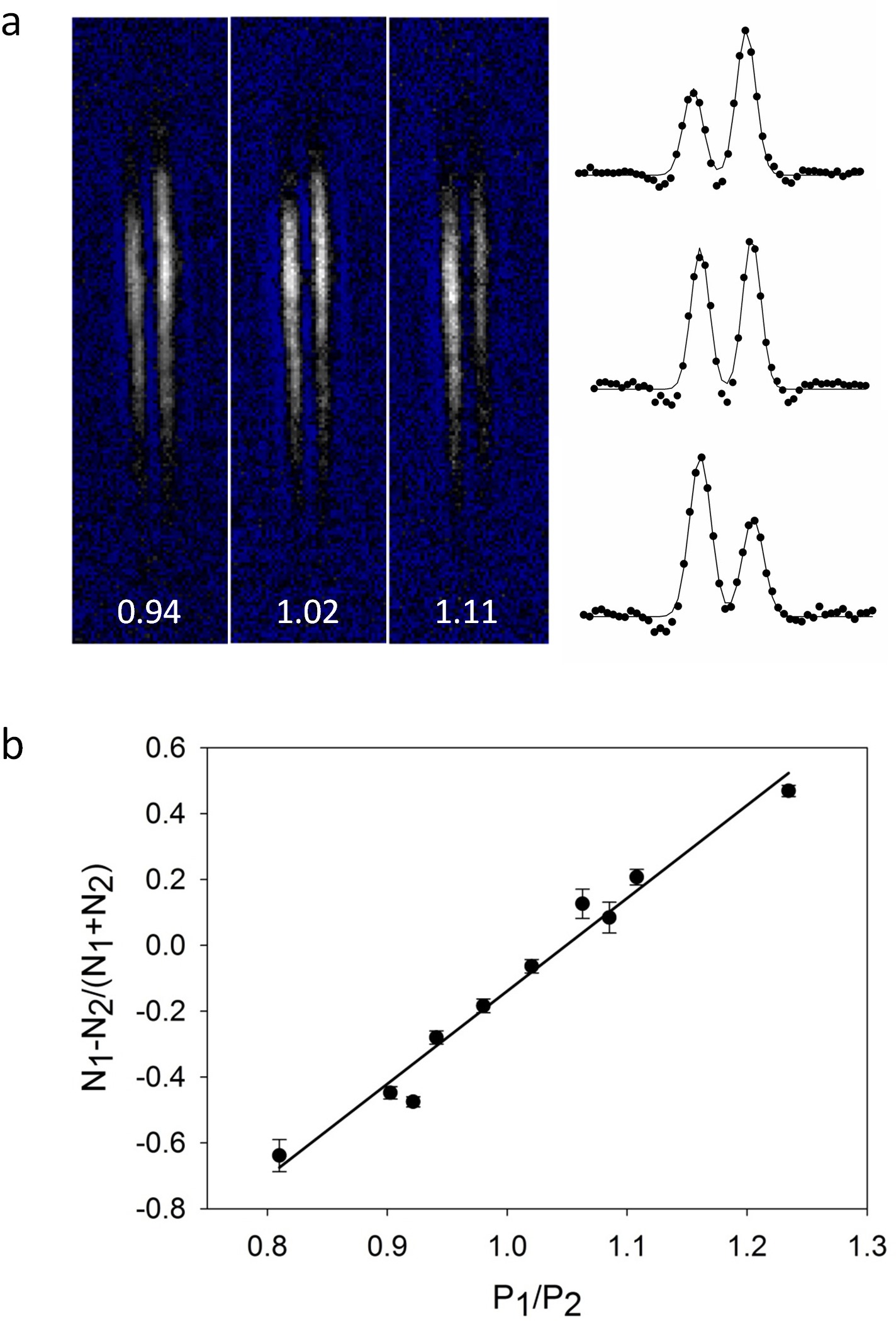}
\caption{\label{Beamsplitter} (a) Absorption images showing that the splitting ratio in the setup used for Fig.\,\ref{Yjunction} can be controlled by altering the relative depth of the potentials in the two arms (labeled on each panel). The dimensions of each image are 86\,$\mu$m\,$\times\,21\,\mu$m.  The three graphs at right show the densities integrated axially. (b) The dependence of the splitting ratio on the ratio of the potential depths of the arms.  The line is a linear fit to the data.}
\end{center}
\end{figure}

\begin{figure}
\begin{center}
\includegraphics[width=14cm]{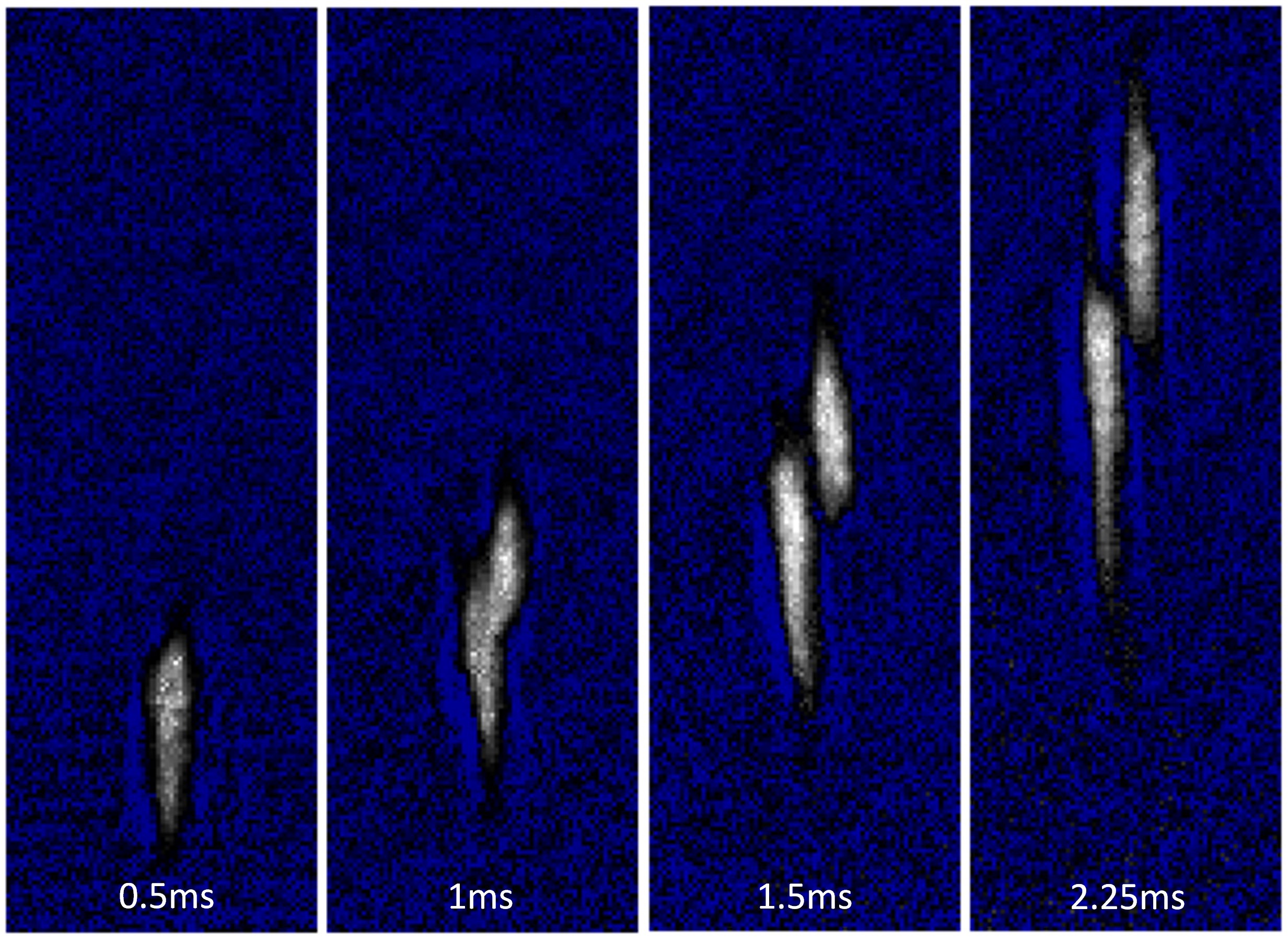}
\caption{\label{switch}Jumping between different intensity ratios in the output arms of the Y-junction realizes a switch, sending the front of the wavepacket down one arm and the back of the wavepacket down the other.  The dimensions of each image are 90\,$\mu$m\,$\times\,30\,\mu$m.  The propagation times shown in each panel are relative to the end of the launching process.}
\end{center}
\end{figure}

A second important circuit component is the Y-junction or beamsplitter.  This circuit element has not been demonstrated previously for guided coherent matter waves.  We found that smoothing of the potential in the junction region was critical to making the Y-junction work, which may explain why splitting of a propagating BEC has not been observed previously.  Further, the painted potential is not subject to the same constraints as potentials produced by static magnetic fields, where a second input waveguide merges with the input waveguide at the splitting point and the transverse confinement at that point is weak \cite{Lesanovsky2006}, which can lead to large perturbations when a condensate is split in that way \cite{Shin2005}.  Figure\,\ref{Yjunction} shows a BEC traveling through a painted Y-junction which smoothly divides it into two pieces.  Altering the relative depths of the potentials in the two arms allows the splitting ratio to be tuned controllably (Fig.\,\ref{Beamsplitter}).  Suddenly changing the intensity ratio from 2:1 to 1:2 while the BEC is in the junction region realizes a switch that sends the front of the wave packet down one arm and the back down the other (Fig.\,\ref{switch}).  It should be possible to extend this scheme to switch multiple input guides between multiple output guides.   This is a simple example of how our approach can realize dynamically reconfigurable circuits.

\begin{figure*}[!]
\begin{center}
\includegraphics[width=15cm]{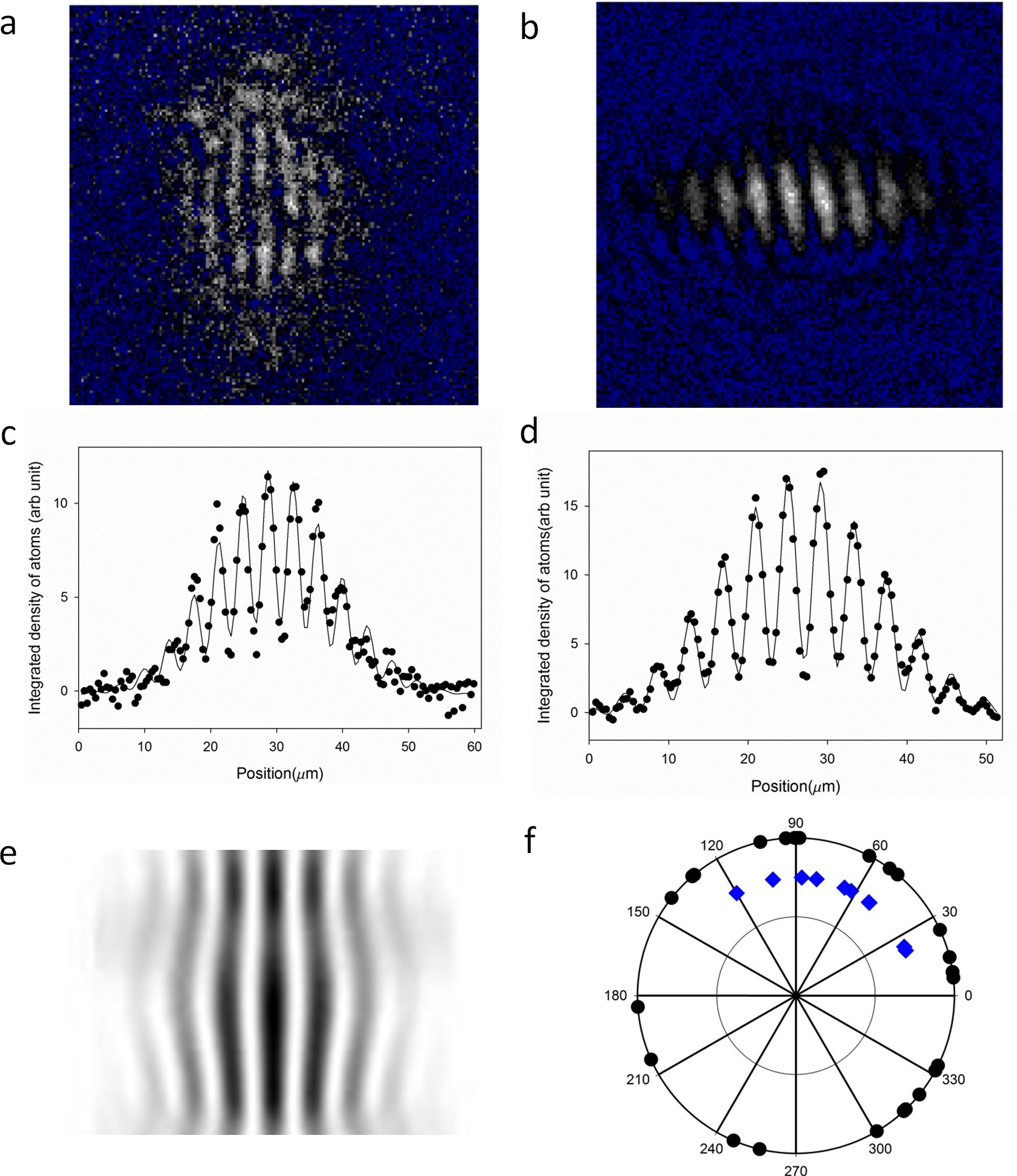}
\caption{\label{phase} Demonstration of a phase coherent Y-junction.  (a) Matter wave fringes formed when the split BEC in the rightmost panel of Fig.\,\ref{Yjunction} is released and allowed to expand for 2.5\,ms.  Image dimensions are $60 \,\mu$m$ \times 60\,\mu$m. (b) Interference fringes formed by BECs created in two separate $12.4\,\mu$m-long potentials in the same locations as the arms of the Y-junction and then allowed to expand for 3\,ms. Image dimensions are $51\,\mu$m$ \times 51\,\mu$m. (c) Fit to the fringes of (a) integrated over the central region of the image. (d) Fit to the fringes of (b) integrated over the central region of the image. (e) GPE simulation of the interference fringes in (a) formed by releasing the BEC after splitting at the Y-junction. (f) Phase of the fringe pattern obtained over several repetitions of the experiment.  Blue diamonds: BEC split in the Y-junction [(a) and (c)], black disks:  BEC created directly in the arm potentials [(b) and (d)].}
\end{center}
\end{figure*}

If a Y-junction is to serve as a useful beamsplitter in a guided atom interferometer it must preserve the coherence of the condensate.  We demonstrated the phase coherence of the division process by allowing the center of the BEC to propagate past the start of the Y-junction for 2\,ms (rightmost panel of Fig.\,\ref{Yjunction}) and then releasing the split condensates to observe interference fringes [Fig.\,\ref{phase}(a)] after $2.5$\,ms of free expansion time.  This procedure realizes a traveling BEC interferometer somewhat analogous to the configuration of a stationary BEC with a time-dependent potential demonstrated in Refs.\,\cite{Shin2004, Schumm2005}.  We note that additional challenges arise when splitting a moving BEC because of the need to avoid transferring significant forward kinetic energy into transverse excitation as the BEC moves through the junction.  This is important because using a propagating BEC provides a way to accomplish the applications discussed in the first paragraph of this paper.  The relative phase of the BECs in the two arms can be determined by fitting a suitable function to the fringe pattern \cite{Shin2004} [Fig.\,\ref{phase}(c)].  The measured phase distribution shown as blue diamonds in Fig.\,\ref{phase}(e) has a standard deviation of $29.7\degree$, much less than the standard deviation of $103.9\degree$ expected for a uniform distribution of random relative phases.  Calculations predict that tunnel coupling between the split condensates should be negligible.  We verified this by forming condensates directly in two long traps separated by the Y-junction arm spacing and analyzing the fringe pattern formed after free expansion [Figs.\,\ref{phase}(b), \ref{phase}(d), and \ref{phase}(f)].  In this case the phase is indeed randomly distributed over all possible angles.  The observed phase coherence therefore must be established by the splitting process.  The measured phase spread is larger than the standard quantum limit $N^{-1/2} \sim 0.6\degree$ for expected for our atom number $N = 10^{4}$.  Understanding the origin of the phase spread and reducing it will be the subject of future work.  The phase coherence time is longer than we can measure at present, but we note that the fundamental limit from phase diffusion in a BEC with $N$ atoms and chemical potential $\mu$, given by $(2 \mu /5h \sqrt{N})^{-1}$ \cite{Shin2004}, is approximately 80\,ms for our conditions.  Close examination of the fringe pattern in Fig.\,\ref{phase}(a) shows that the fringes are slightly concave with respect to the central fringe.  GPE simulations confirm that this is due to excitations produced in the splitting process [Fig.\,\ref{phase}(e)].  Once these excitations are reduced, for example by fine tuning the shape of the potential at the division point, it will be interesting to connect two such Y-junctions back-to-back to form a waveguide Mach-Zehnder interferometer.

\section{Summary and Outlook}
We have shown here how to create simple circuits for coherent atomic matter waves.  The system, which uses time-averaged optical dipole potentials, is able to launch matter waves from a BEC into a waveguide at a desired velocity, and then propagate the matter waves almost single-mode around bends connecting straight waveguide sections.  It can also switch propagating matter waves, and divide the moving matter waves phase-coherently to realize a simple atom interferometer.  This demonstration of the basic matter wave circuit elements opens the door to the creation of complex and dynamic matter wave circuits.  In particular, the scalable circuit technology reported here is well-suited to creating guided atom interferometers \cite{Cronin2009} and to both realizing atomtronic devices \cite{Seaman2007} and wiring them together to create functionality.

A next step is determining the optimum shape for bends and Y-junctions to minimize excitation out of the ground state, and then implementing these geometries using the technology described here.  Looking to the future, it will be possible to add further circuit functionality by painting potential structures on waveguides to act as partial (or total) reflectors for matter waves, extending the analogy with integrated optics.  One can of course use a simple potential barrier with appropriate height to reflect (or transmit) matter waves.  However this form of mirror/beamsplitter has the disadvantage for manipulating wavepackets that its reflectance and transmittance vary rapidly with de Broglie wavelength.  One solution to this problem is well-known from traditional optics:  replace the single potential barrier by a periodic array of weaker potentials.  This is the basic principle of broadband multilayer optical coatings and of distributed Bragg reflectors.  Such reflectors have been realized for BECs by imposing an optical lattice on a collimated red-detuned laser beam serving as a guide \cite{Fabre2011}.  In our case we could form a waveguide Bragg reflector by painting a periodic modulation in the waveguide potential.  Simulations show that with as few as six wells the transmission is almost flat for incident energies from 100\% to 200\% of the modulation depth.   A further next step is to combine two such reflectors on the same guide to realize a matter wave cavity.  This atom-optical configuration has been studied theoretically for almost twenty years \cite{Wilkens1993, Paul2005}, beginning with a proposal to use such a cavity as an extremely narrow-band velocity filter \cite{Wilkens1993}, an application which remains relevant today.  In addition, since atoms, unlike photons, can interact strongly it may be possible to operate such a cavity in a nonlinear regime.  In the case of strong nonlinearity it is predicted \cite{Carusotto2001} that an atom blockade effect will be seen, where only one atom can occupy the cavity at a time and the de Broglie wave leaking out of the cavity exhibits non-classical statistics.  Additional possible future directions to study with atomtronics using the technology presented here include phase squeezing in matter wave beamsplitters \cite{Jo2007, Grond2009}, quantized conductance of atoms \cite{Thywissen1999, Krinner2015}, and creating complex superfluid circuits containing devices such as atom-SQUIDs \cite{Wright2013, Ryu2013, Eckel2014}.

\ack{This work was supported by the U.S. Department of Energy through the LANL/LDRD Program.  We acknowledge inspiring conversations with Eddy Timmermans.}

\end{document}